\documentclass[preprint,journal]{vgtc}       

\ifpdf
  \pdfoutput=1\relax                   
  \usepackage{graphicx}                
  \DeclareGraphicsExtensions{.pdf,.png,.jpg,.jpeg} 
\else
  \usepackage{graphicx}                
  \DeclareGraphicsExtensions{.eps}     
\fi%

\graphicspath{{figures/}{pictures/}{images/}{./}} 

\usepackage{microtype}                 
\PassOptionsToPackage{warn}{textcomp}  
\usepackage{textcomp}                  
\usepackage{mathptmx}                  
\usepackage{times}  
\usepackage[normalem]{ulem} 

\usepackage{cite}                      
\usepackage{tabu}                      
\usepackage{booktabs}                  
\usepackage{hyperref}

\usepackage{amsmath,amsthm,amssymb,hyperref}
\usepackage[linesnumbered,ruled,vlined]{algorithm2e}

\SetCommentSty{mycommfont}
\SetKwProg{Df}{DeviceFunction}{:}{}
\SetKwProg{Kn}{Kernel}{:}{}
\SetKwProg{Fn}{Define}{:}{}
\SetKwRepeat{Dw}{do}{while}
\SetKwFor{For}{for }{}{}
\usepackage{float}
\usepackage{mathtools}
\DeclarePairedDelimiter\ceil{\lceil}{\rceil}
\DeclarePairedDelimiter\floor{\lfloor}{\rfloor}
\usepackage{amsfonts}
\usepackage{mathtools}
\usepackage{tikz}
\usepackage{pgfplots}
\pgfplotsset{compat=1.17}
\usepackage{url}
\usepackage{chemformula}
\usepackage{enumitem}
\usepackage[utf8x]{inputenc}

\onlineid{1496}

\vgtccategory{Research}
\vgtcpapertype{algorithm/technique}

\title{Level Set Restricted Voronoi Tessellation\\ for Large scale Spatial Statistical Analysis}

\author{Tyson Neuroth, Martin Rieth, Konduri Aditya, Myoungkyu Lee, Jacqueline H Chen, and Kwan-Liu Ma}
\authorfooter{
\item
 Tyson Neuroth and Kwan-Liu Ma are with University of California, Davis. E-mail: taneuroth@ucdavis.edu, ma@cs.ucdavis.edu.
 \item
 Martin Rieth and Jacqueline H Chen are with Sandia National Laboratories. E-mail: mrieth@sandia.gov, jhchen@sandia.gov.
 \item
 Konduri Aditya is with India Institute of Science. E-mail: konduriadi@iisc.ac.in
  \item
Myoungkyu Lee is with University of Alabama. E-mail: mklee@ua.edu.
}

\shortauthortitle{Neuroth \MakeLowercase{\textit{et al.}}: LSRVT for Stat. Analysis}

\abstract{

Spatial statistical analysis of multivariate volumetric data can be challenging due to scale, complexity, and occlusion. Advances in topological segmentation, feature extraction, and statistical summarization have helped overcome the challenges. This work introduces a new spatial statistical decomposition method based on level sets, connected components, and a novel variation of the restricted centroidal Voronoi tessellation that is better suited for spatial statistical decomposition and parallel efficiency. The resulting data structures organize features into a coherent nested hierarchy to support flexible and efficient out-of-core region-of-interest extraction. Next, we provide an efficient parallel implementation. Finally, an interactive visualization system based on this approach is designed and then applied to turbulent combustion data. The combined approach enables an interactive spatial statistical analysis workflow for large-scale data with a top-down approach through multiple-levels-of-detail that links phase space statistics with spatial features.
} 


\teaser{
  \centering
  \includegraphics[width=\linewidth]{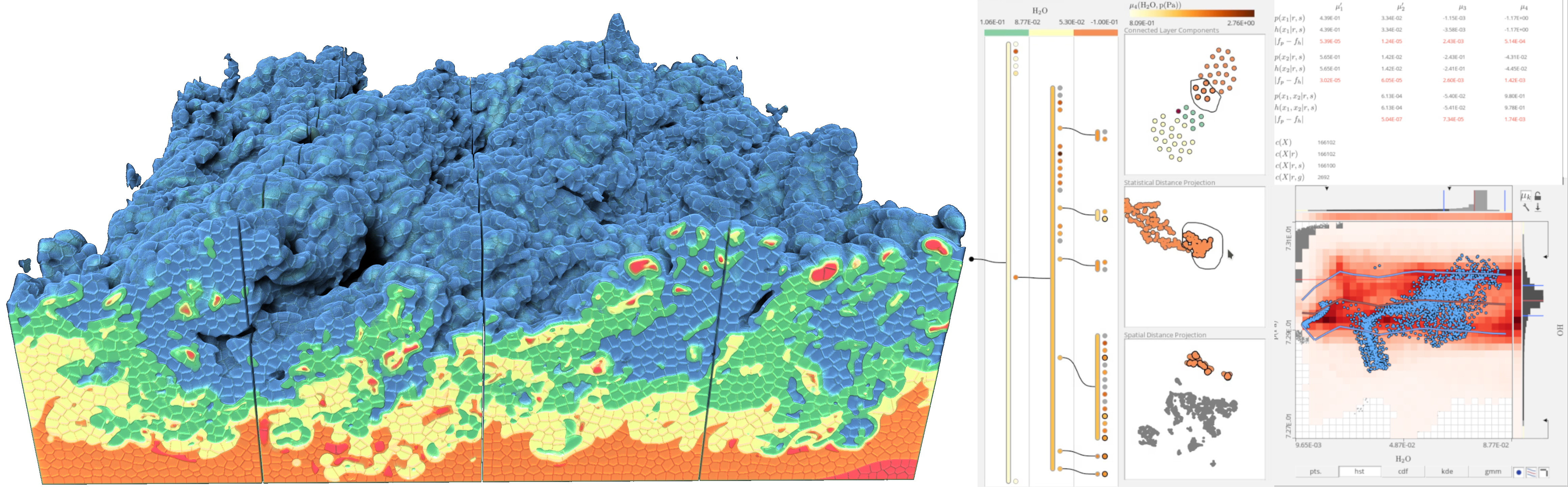}
  \vspace{-0.25in}
  \caption{Left to right illustrates steps of our methodology. We decompose volume data by novel geodesically based discrete centroidal Voronoi tessellations restricted by sequences of level sets. Then we aggregate statics within the resulting hierarchically structured segments. Finally, the data is reorganized according the hierarchical structure, and used for interactive spatial statistical analysis of large data at a reasonable cost.}
  \label{fig:teaser}
}

\vgtcinsertpkg

\begin{document}


\firstsection{Introduction}
\maketitle

Simulation scientists continue to push computational limits 
for the modeling of complex spatial systems. These simulations can generate extreme-scale data with high spatial and temporal fidelity, but I/O bottlenecks and storage constraints place limitations on how much of this data can be saved and analyzed post hoc. This problem motivates the development of in situ data analysis and processing workflows with the goal of reducing I/O while retaining crucial information.

One way to approach this problem is spatial statistical aggregation. Such approaches usually involve segmentation or decomposition in the spatial domain, and then statistical aggregation within the resulting segments. Several methods have been developed for spatial segmentation of statistical aggregation, such as blockwise decomposition, Voronoi decomposition, and voxel clustering. These methods have demonstrated value; however, it remains a challenge to organize such a tessellation around dynamic flow features, typically defined by surfaces. 

To address this problem, we segment the spatial domain dynamically using level set restricted centroidal 
Voronoi tessellation (CVT). The CVT may also be weighted to vary the spatial fidelity based on a spatial saliency function. Since the decomposition is aligned to the level sets, this also provides a hierarchical segmentation, from isbands (voxels between pairs of isosurfaces or contours) to connected components within isobands, to the centroidal tessellations within the connected components. The statistical aggregations may also be merged over Voronoi cells and throughout the hierarchy. For example, one may view statistics of the Voronoi cells individually, or locally merge Voronoi cells to obtain aggregations over multi-segment volumetric patches that are aligned to the surfaces of interest, or merge all Voronoi cells within a connected component or entire isoband.

For practical purposes, the implementation must be scalable and efficient. Efficient algorithms have been developed for computing CVTs; however, the focus of these methods has been on different application areas, such as re-meshing and construction of computation grids. These other applications have requirements that Voronoi must conform to, and are typically defined through precise geometry. For statistical aggregation, however, the requirements may be relaxed, since heterogeneity in region size but conflict with the goal of generating a conforming mesh in complex geometries. Meanwhile, the benefit of the discrete tessellation is that it can be done efficiently in parallel on GPUs and it fits the analysis workflow whereby voxel elements are sub-selected discretely for visualization. Since our requirements differ from the typical use case, and because we are aiming for high performance, we implemented our own discrete method for a non-conforming level set restricted tessellation that favors volumetric and geodesic regularity.

{   
Additionally, interacting with the data in this form is difficult without a supporting visualization system. As scientists put great preparation into planning an expensive, large-scale simulation run, it is important to also adequately prepare the configuration of the in situ workflow so that it can be relied on without concern. For example, with our approach, one must choose level sets, overall Voronoi site density, and the optional saliency function for concentrating site density. For statistical aggregation, there are several possible models for distribution functions, each with different possible parameters that should be tuned appropriately. Among models and their parameters, there are also tradeoffs affecting performance, storage costs, fidelity, and other measures of quality. Thus a working visualization system is a valuable tool to have before one attempts to apply the in situ methods. Our visualization system is able to help with this, since it can be used on existing data where the methods can be tested and the aggregates can be compared with the raw data to get an idea what information may be lost or misrepresented by the aggregates. Figure~\ref{fig:teaser} shows the system interface.



Our primary contributions are
\begin{itemize}[noitemsep,topsep=0pt]
    \item a parallel level set restricted CVT algorithm and implementation,
    \item a supporting visualization system, and
    \item an overall approach to visualizing large scale multivariate spatial statistical information.
\end{itemize} }

We apply our work to simulation data of turbulent combustion and wall-bounded turbulence, and we evaluate the performance and scalability on the Summit supercomputer at the Oak Ridge National Laboratory.

\section{Related Work}
\label{sec:related} 

This work is inspired by past success using spatial statistical aggregation to reduce and analyze large simulation data. Chadhuri et al. introduced scalable distributed algorithms for computing histogram-based spatial summaries of large 3D mesh data~\cite{6378985}. Lu et al. introduced compact histograms representations based on bit-strings and space-filling curves~\cite{7348071}. Histograms have been used to summarize  space phase distributions of statistical super-particles in particle-in-cell simulation data~\cite{neuroth2015scalable,kress2018binning}. For simulation monitoring and anomaly detection in transonic jet engine simulations, distribution functions were used to quickly and compactly represent flow conditions, so that they could be monitored in real-time for human in the loop stall detection~\cite{7539561}. Ye et al. decomposed the domain of combustion simulations into blocks, then computed spatial statistics for querying Lagrangian particles based on histogram and coupled data structure~\cite{7874311}. Many more related methods have been introduced. Gaussian mixture models (GMMs) and copula functions have provided small footprint models of joint statistics~\cite{Dutta17PacVis,8017601}. Additional work has focused on the recovery of the spatial distributions~\cite{Wang17PacVis}. In terms of spatial segmentation for statistical aggregation, further approaches have been applied, including kd-trees, and simple linear iterative clustering~\cite{Dutta17PacVis}.

The aforementioned work~\cite{neuroth2015scalable, 7539561, 7874311}, each included interactive visualization systems for local spatial statistical analysis with distribution functions computed over local regions, but were each designed with different goals or for application domains. Neuroth et al. introduced a design which relied on a 2D decomposition for Tokamak data~\cite{neuroth2015scalable}; where symmetry justifies radial integration to project the data onto the 2D plane. They used a static Voronoi tessellation. Like our system, their system also supported selecting and merging the statistics in different spatial regions, which could then be analyzed through linked plots. However, they only had static plot parameters. Ye et al.\cite{7874311} utilized a Cartesian spatial decomposition of a 3D domain for combustion data. This has an efficiency advantage but cannot support feature-aligned spatial statistical analysis. Dutta et al. probed spatial statistics around the surface of engine blades~\cite{7539561}. Their system provided human-in-the-loop anomaly detection utilizing a streaming timeline visualization. While these systems may be considered state of the art in their respective ways, we are not aware of any approach and supporting system which can provide the type of interactive statistical spatial analysis around features that our system supports.

Besides tailored visualization systems for large-scale spatial statistical analysis, there are state-of-the-art tools for accommodating many general visualization tasks, both in situ and at scale. Paraview~\cite{ahrens2005paraview} is one such platform, and Visit~\cite{childs2012visit} is another such tool. These platforms are indispensable for supporting large-scale visualization and analysis across an extremely wide range of dataset formats and scientific domains. However, while they can also be heavily customized and used in integrated workflows, they do not have the capability by default to replicate the interactivity and features of our system for linked spatial statistical analysis. Nevertheless, alternate or similar workflows could be used, perhaps with an integrated approach incorporating multiple tools. The visualization toolkit (VTK)~\cite{schroeder2000visualizing} pipelines would work well to support the data processing pipeline of the Voronoi processing and statistical aggregation. Our algorithms, which were implemented in CUDA~\cite{nickolls2008scalable} (NVIDIA's GPU computing API) and Thrust~\cite{bell2012thrust} (a high-level interface for data-parallel computing), could also be implemented using VTK-m~\cite{moreland2016vtk} (a toolkit of parallel scientific visualization algorithms), which is similar in its model to Thrust and CUDA.


Besides approaches based on statistical modeling, there has been a great deal of work done on sub-sampling data from large complex domains while preserving surface and topological features in flows~\cite{adams2007adaptively, woodring2011situ,6171182, zhang2008adaptive}. Topological methods for visualization, in general, are too numerous to mention. Heine et al. provided an in-depth survey~\cite{heine2016survey}. Carr et al. introduced efficient parallel algorithms to compute contour trees~\cite{7874312,carr2019scalable}. Bremer et al. proposed using merge tree as a way to semgement combustion data into hierarchical features that can be summarzied through statistics and explored efficiently through their graphs~\cite{bremer2009topological}. Another great idea that has pushed us forward is the work on summarizing large time-varying simulations through the use of dynamic nested tracking graphs~\cite{lukasczyk2019dynamic} in a coupled interface that links the topological features to in situ rendered images within a cinema database~\cite{7013022}. In addition, recent work has offered a generalization of isosurfaces to multi-fields, which inspires an entirely new approach towards level set and topological based analysis in multivariate data~\cite{8453863}. These are just a few examples, as the area of topological segmentation is a rich and active area of research. Our method utilizing a restricted centroidal Voronoi tessellation (CVT) is designed to compliment these methods, by offering an additional capability to decompose the topological segments further without losing the benefits. 



Voronoi tessellations, also called Voronoi diagrams or Voronoi decompositions, are among the oldest and most fundamental methods in computational geometry and related fields~\cite{aurenhammer1991voronoi,du2002grid}. The CVT is also useful for glyph placement and stippling~\cite{du2004centroidal, gortler2019stippling}. The centroidal Voronoi tessellation, where each Voronoi site is at the centroid of its region, is commonly computed using Lloyd's Algorithm~\cite{lu2016statistical}. A centroidal Voronoi generates a segmentation with more homogeneous region sizes and shapes, which is more favorable for spatial statistical analysis. Our contribution is a CVT algorithm that is designed for spatial statistical analysis and summarization of flow features, which are complex in shape and encapsulated by level sets. 

{
CVTs with restrictions/constraints have been introduced~\cite{tournois20102d}, however this has been made possible in complex shapes by concentrating more Voronoi sites around the sharp/complex features, often with the goal of geometric modelling and surface simplification or reconstruction. For example, Amenta et al. introduced PowerCrust, which is able to generate a `watertight' boundary 3D polyhedral solid, with theoretical guarantees on quality. VoroCrust expanded on this line of work with additional provable guarantees under more general conditions~\cite{abdelkader2020vorocrust}. Once a `conforming' decomposition of the boundary has been constructed, the interior could then be decomposed into a centroidal tessellation using existing algorithms. Voronoi tessellations on manifolds and within non-Euclidean spaces based on geodesic distances have been introduced as well~\cite{10.1145/1839778.1839795,WANG201551}, where the geodesic represents the shortest path between two points on the manifold. The manifold could be surfaces in 3D space or higher dimensional, \textit{e.g.} the 3+1D hyperbolic space in general relativity. Note the geodesics in our method are not geodesics on a manifold, but rather graph theoretic geodesics representing shortest paths obeying the level set restrictions.}

{
\subsection{Motivation, Research Challenge and Novelty}

\textbf{Restrictions and Homogeneity}. In Section~\ref{sec:related}, we discussed the state of the art in creating Voronoi tessellations around complex restrictions. The state of the art methods work by building a conforming crust around the boundaries. But in building the crust so that it is conforming,  we must concentrate more sites around sharp features, and thus we will sacrifice homogeneity of the Voronoi regions on the boundaries. But the regions on the boundaries are those of primary interest for spatial statistical analysis. This is why we relax conditions required for surface reconstruction, and instead focus on homogeneity of the region sizes. However, this creates challenges for a restricted CVT algorithm since individual regions will wrap around the flow features, and so the normal centroid is no longer valid since it may not even be within the region, and the centroidal updates could carry a Voronoi site through a boundary/restriction. Therefor, we create a novel restricted CVT algorithm based on graph geodesic distances, where the graph is constructed within the volume for each voxel to have a shortest path around the restriction boundaries to the graph geodesically nearest Voronoi sites. Besides CVTs, homogeneity guided decompositions for spatial statistics have been approach using simple linear iterative clustering (SLIC)~\cite{Dutta17PacVis}, however, to our knowledge, an algorithm for restricted SLIC is not available, and since the spatial component of SLIC is based on Euclidean distance, a restricted version also could not reliably meet complex boundary restrictions while maintaining spatial homogeneity of region sizes.

\textbf{Scalability and Computational Performance}. Since the domain is evolving, and one application of the decomposition is for in situ processing, it is important that our algorithm is efficient, and exploits the heterogeneous parallelism of modern super computers, including distributed data parallelism, and GPGPU parallelism. Comparisons between our algorithm and others are difficult since to our knowledge, no other algorithm computes the same form of CVT as ours. However, algorithms have been proposed for standard CVTs on 2D planes and surfaces which exploit GPU parallelism~\cite{rong2010gpu}, and in 3D parallel clipped CVTs~\cite{yan2010efficient}. A general data distributed parallelism for CVTs was intruduced by Starinshak et al.~\cite{starinshak2014new}, which also handles restrictions only by clipping. However, clipping will leave randomly fragmented regions along the boundaries depending on the data. 

\begin{figure}[h]
  \centering
  \includegraphics[width=\linewidth]{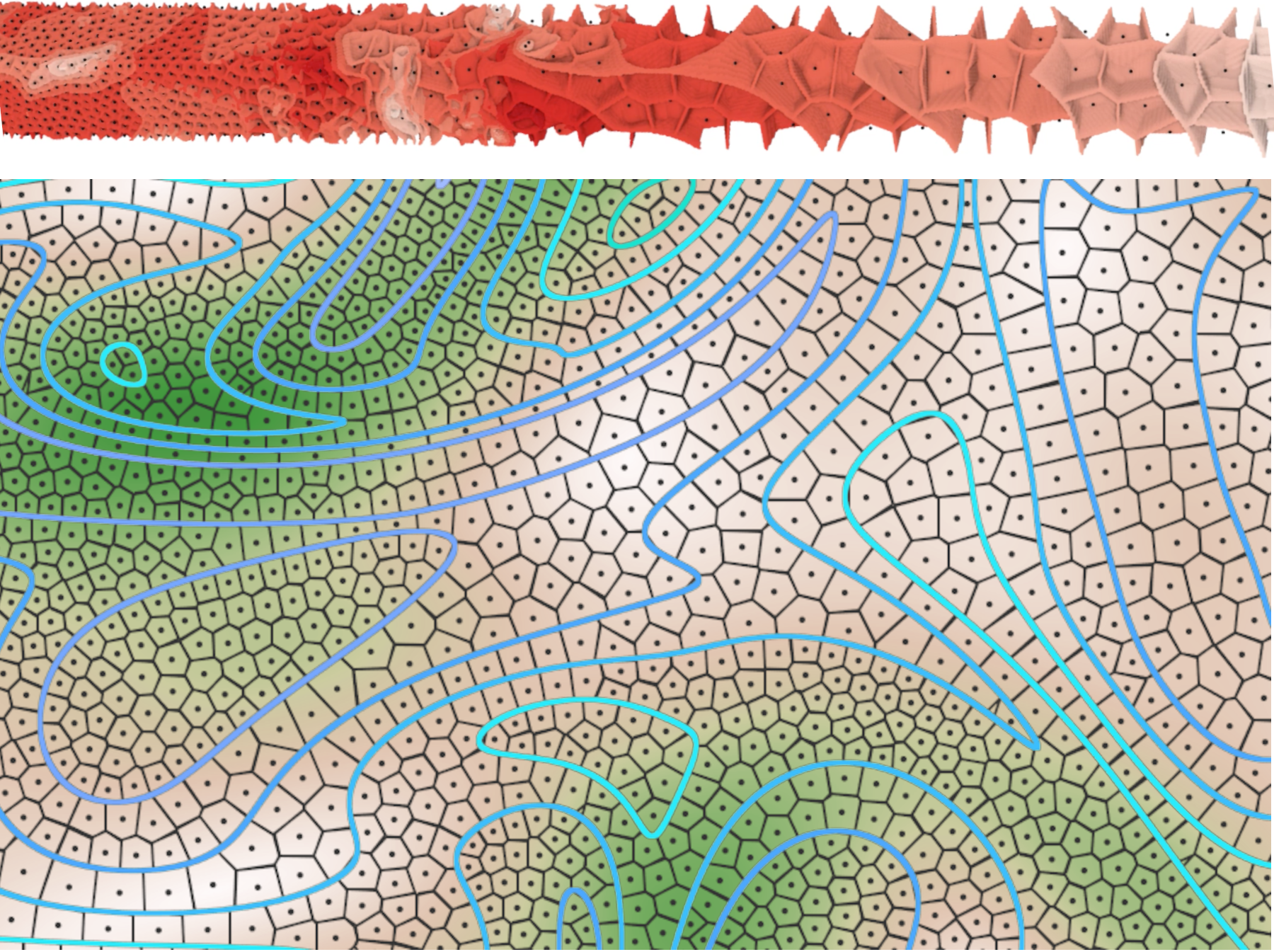}
  \vspace{-0.3in}
  \caption{A 2D case of the weighted centroidal Voronoi tessellation, where contours of one scalar field (shades of blue) impose the restrictions, and the second field (brown to green) controls the site density.}
  \vspace{-6mm}
  \label{fig:weighted2d}
\end{figure}

\section{Methods}
\label{sec:methods}

First, we introduce a local statistical aggregation approach based on a discrete geodesically based level set restricted CVTs. It is geodesically based in a graph theoretical sense, with voxels as nodes and the geodesics being the shortest restricted paths from each voxel to its nearest Voronoi site. We then present a prototype system that utilizes the method for hierarchical spatial statistical analysis.

\begin{figure}[ht]
  \centering
  \includegraphics[width=0.9\linewidth]{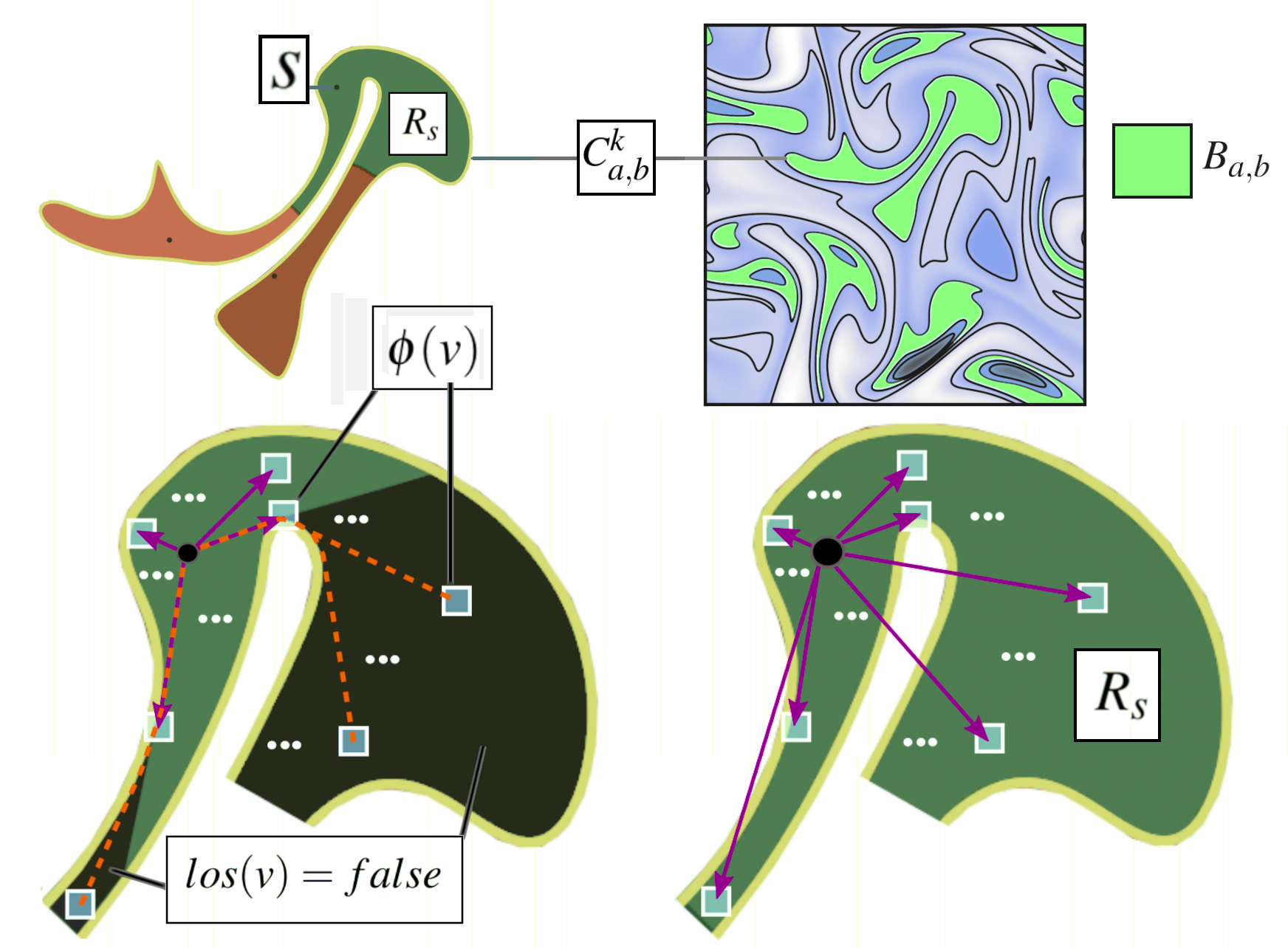}
  \vspace{-0.15in}
  \caption{Upper right) Domain decomposed by isocontours into isobands. One isoband $B_{a,b}$ is highlighted in bright green. Upper left) One connected component of the isoband $C_{a,b}^k$ along with 3 Voronoi sites/regions. Lower right) Voronoi region $R_s$ and illustration of standard averaging of the vectors for all voxels. Lower left)  The mass of non \textit{line-of-site} voxel $v$ is propagated to $\phi(v)$, the LOS voxel on its geodesic path.}
  \vspace{-5mm}
  \label{fig:process}
\end{figure}

\subsection{Level Set Restricted Voronoi Tessellation}
\textbf{Level Sets, Isobands, and Connected Components}. We define a set of disjoint discrete isobands where each isobands $B_{a,b}$ is the set of voxels which have values between scalar values $a,b$, \textit{i.e.} $B_{a,b} = \{ v \in V \mid a < f(v) \leq b \}$. The isobands may be comprised of multiple spatially disconnected components, $C_{a,b}^k$, such that $B_{a,b} = C_{a,b}^0 \cup C_{a,b}^1 \cup ... \cup C_{a,b}^m$, where no voxel in $C_{a,b}^i$ is adjacent to any voxel in $C_{a,b}^j$ for any pair $(i,j)$. Each of these components is a bounded region which is to be tessellated by a CVT. 

\textbf{Voronoi Sites and Regions}. We define the Voronoi site $s$ is a defined by a point $\mathbf{x}_s$, and its Voronoi region restricted to the component $C$ as
\begin{equation}
\label{eq:1}
R_s = \{ v \in C \mid s = min_{u \in S_C} gd(\mathbf{x}_v,\mathbf{x}_{u}) \}
\end{equation}
  
\noindent
where $S_C$ is the set of Voronoi sites in component $C$, and $gd(\mathbf{x}_v,\mathbf{x}_{u})$ is the graph geodesic distance from voxel $v$ to Voronoi site $u$, where each voxel may be node in the graph, and no edges in the graph cross any boundaries, so that the distance $gd(\mathbf{x}_v,\mathbf{x}_{u})$ is the shortest restricted path from the voxel to the Voronoi site. 

\textbf{Centroids or Centers of Mass}. A standard discrete CVT in a volume is a Voronoi tessellation where each Voronoi site point is at the centroid or center of mass of the region it encapsulates, where center of mass is defined as $\frac{1}{m(R_s)}\sum_{v \in R_s} m_v\mathbf{x}_v$, where $R_s$ is the region of the site $s$, $v \in R_s$ are the encapsulated voxels, $m_v$ is the weighting function evaluated at volume element $v$, $\mathbf{x}_v$ is the position vector of $v$, and $m(R_s)$ is the net mass for all $v\in R_s$. In the unweighted version, $(\forall v)[m_v = 1]$. However, by this definition, the center of mass of a Voronoi region could be outside the boundaries, which would make the centroidal distribution impossible, create disconnected Voronoi regions, and break Loyde's algorithm (which repeatedly moves each site to the center of mass). To avoid escaping its component, we can move the site towards the centroid only until it runs into the boundary, but the sites can still remain stuck moving into the obstacles instead of around them. Instead we compute a weighted centroid of only \textit{line-of-site} (LOS) voxels. The LOS voxels are those for which a strait line between them and their Voronoi site does not intersect a boundary. To account for voxels which are not LOS, we propagate their mass to the line of site nodes in their geodesic path back to their Voronoi site. This leads us to a new definition of the centroid $\mathbf{x}_c$ , where instead of $\mathbf{x}_c = \frac{1}{m(R_s)} \sum_{v\in{R_s}} m_v \mathbf{p}_v$, we define,
\vspace{-1mm}
\begin{equation}
\label{eq:2}
\mathbf{x}_c'= 
 \frac{1}{m(R_s)}\sum_{v\in R_s} m_v \mathbf{x}_{\phi(v)}\\ 
 \text{ s.t. } \phi(v) = v \textbf{ \textit{if} } los(v,s) \textbf{ \textit{else} } \phi(src(v))
 \vspace{-1mm}
\end{equation}
 \noindent
 where $los(v,s) = true$ \textbf{\textit{if}} $v$ is line of site to $s$ \textbf{\textit{else}} $false$. Figure~\ref{fig:weighted2d} shows weighted 2D and 3D examples of our LSRCVT.

\begin{figure*}[t]
  \centering
  \includegraphics[width=0.99\linewidth]{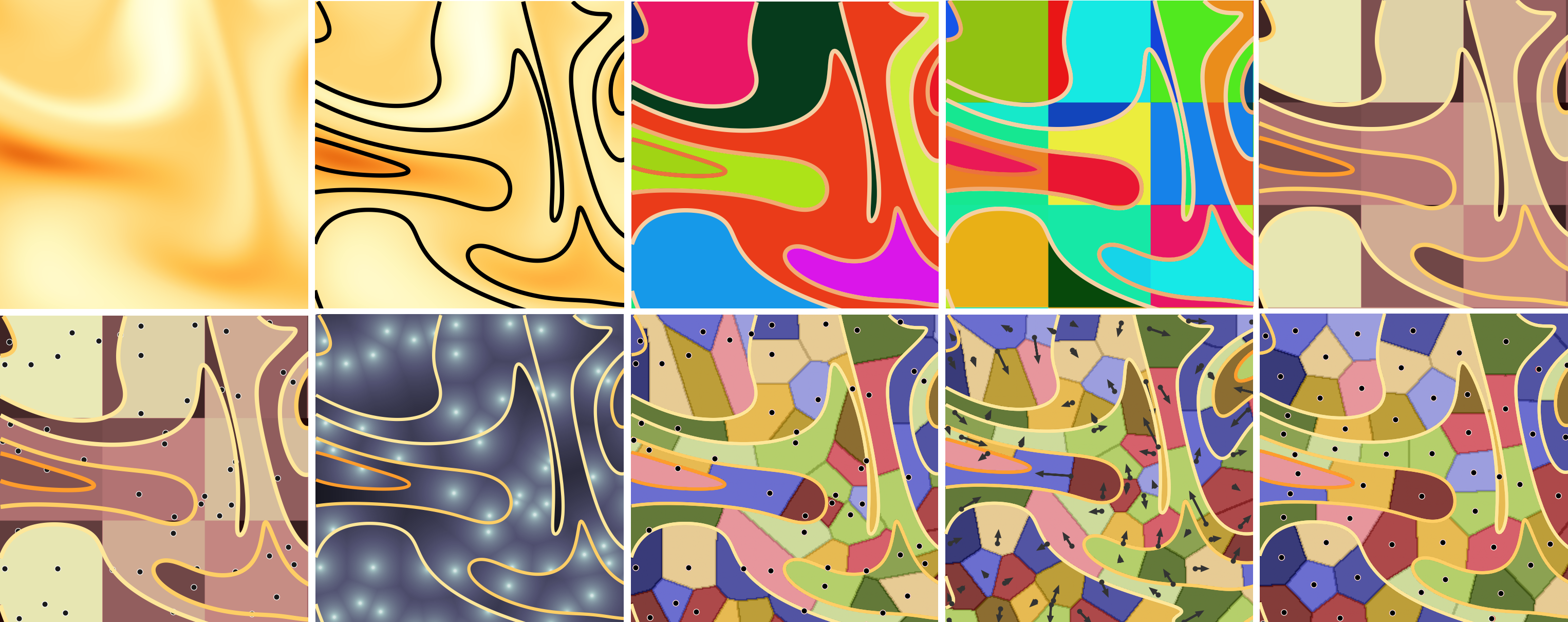}
  \caption{High level illustration of our CVT algorithm. Left to right, top to bottom: The scalar field, decomposing into layers based on a set of isosurfaces/contours, connected component labelling, component/block decomposition for stratified component sampling, heat map showing the mass of each segment which determines how many Voronoi sites to place in each segment, the seed Voronoi sites, geodesic distances from each voxel to its geodesically nearest site, the initial Voronoi tesselation, a centroidal update, and lastly, the tesselation after multiple centroidal updates.}
  \vspace{-5mm}
  \label{fig:steps}
\end{figure*}


\begin{algorithm}
{\scriptsize
    \DontPrintSemicolon
    \Fn{LRCVT() } {
        StratifiedRandomSiteDistribution() \;
        \Dw{ \textbf{not} converged() }
        {
            RestrictedGeodesicVoronoiDecomposition()\;
            RestrictedGeodesicallyWeightedUpdate()\;
        }
        GeodesicVoronoiDecomposition()\;
    }
    \label{alg:1}
    \caption{Level Set Restricted CVT}
    }
\end{algorithm}

\subsection{Parallel Algorithm and Implementation}

Our algorithm is based on Loyde's algorithm~\cite{du1999centroidal}, which repeatedly computes the Voronoi tessellation (VT) and then updates each Voronoi site to the centroid of its region as shown in Algorithm 1. The difference is we use the geodesic distances defined in Equation~\ref{eq:1}, and illustrated in Figure~\ref{fig:process}, to compute the VT, and we update the site positions based on $\mathbf{x}_c$ as defined in Equation~\ref{eq:2}. The full process is illustrated in Figure~\ref{fig:steps}.

For distributed parallelism, each rank computes a LSRCVT for one block of data independently. The result is that the CVT is aligned to both the level set restrictions and the block boundaries, as can be seen in the large scale tessellation in Figure~\ref{fig:teaser}. The benefits and limitations are discussed in Section~\ref{sec:discussion}. Within each block, GPU parallelism is used to compute the CVT. The code is written in C++ using MPI for distributed parallelism, and Thrust~\cite{bell2012thrust} and CUDA~\cite{nickolls2008scalable} for GPU parallelism. Due to space limitations, the full implementation details are included in the supplemental material.

\subsubsection{Stratified Random Initial Site Distribution.} 

The initial Voronoi sites are generated to cover the set of connected components with a target number of sites $n_c$ in each component chosen based on its net mass and a \textit{site-density} parameter $\alpha$ using the formula 

\vspace{-1mm}
\begin{equation}
\label{eq:nc}
n_c = max( \ceil{ \frac{\alpha}{m(C, \gamma)} \sum_{v\in C}m_v^{\gamma} }, 1 ).
\vspace{-1mm}
\end{equation}

\noindent
Where $\gamma$ is a parameter, the \textit{attraction}, that reduces or increases the effect of the weighting, $m(C,\gamma)=\sum_{v\in C}m(v)^{\gamma}$, and $m(v)$ is the voxel's weight. We also use a block-wise spatial stratification, where each component is segmented into pieces and then the $n_c$ sites are distributed fairly over those pieces. This is illustrated in Figure~\ref{fig:steps} row 1 (r1), columns 3,4, and 5, where the components (c3) are split into blocks (c4), the net mass of each component block is computed (c5), and then sites are placed based on the net mass~\ref{eq:nc} (r2,c1).

The process is done in parallel on the GPU. First we compute the net component masses $m(C, \gamma)$ and the net component block masses $m(b_i,\gamma)$. Next, the $n_c$ sites are distributed by first seeding $\floor{n_c \frac{m(b_i, \gamma)}{m(C,\gamma)}}$ sites in each block, and then placing the remaining sites to the most underrepresented blocks based on the equation,

\vspace{-1mm}
\begin{equation}
ur(b_i)=1.0 - n_{b_i} + \floor{n_c \frac{m(b_i, \gamma)}{m(C,\gamma)}}
\end{equation}

\vspace{0.07in}

\subsubsection{Discrete Restricted Geodesic Voronoi Tessellation.}
\label{sec:VT}

The Voronoi tessellation, defined by Equation~\ref{eq:1}, is computed using a GPU parallel region growing approach emanating from the Voronoi sites. The voxels communicate with their neighbors to determine if there is a possible shorter path, and the process continues until all voxels have paths and no voxel can find a shorter one.

The paths $P_v$ are represented recursively through a map $src(v) : V \longrightarrow V$, and can be constructed by the formula $P_v = ( v, src(v), src(src(v)), ..., \phi(v), s)$, where $\phi(v)$ (defined in Equation~\ref{eq:2}) is the node in the path which is in \textit{line-of-site} of $s$. Each voxel also has an 1-byte bit-field, $state_v$. One bit indicates $los(v,site(v))$ (whether v is \textit{line-of-site} of its site), another indicates $active(v)$ (whether $v$ has a current path), and another indicates $node(v)$ (whether $v$ is a valid node).

In the initial step, the nearest volume element to each site is classified and activated. Then, repeatedly until convergence, for each voxel $v$ in parallel, we ask each other adjacent voxel $w$ if it is active, and if it is, then their are two possible paths for $v$ back to $s$: (1) the path $(v,P_w)$, which is has the distance $||\mathbf{x}_v- \mathbf{x}_w||_2 + d[w]$, or (2) the path $(v, P_{src(w)})$, which has distance $||\mathbf{x}_v - \mathbf{x}_{src(w)}||_2 + d[src(w)]$. The former path (2) might intersect a boundary, so a ray must be cast from $v$ to $src(w)$ to check for a collision. If $v$ finds a shorter path based on the best of these possible choices, then it records the distance and new $src(v)$, but the update is delayed until a next kernel invocation to avoid race conditions. In order to determine which voxels are LOS to their Voronoi sites, in a initial stage of the above procedure, we only allow $s$ to be active nodes so that all classified voxels are \textit{line-of-site}. Then we activate the rest of the voxels as nodes and perform the procedure repeatedly until voxels have paths and cease to find a better path. 

Figure~\ref{fig:steps} row 2 column 3 shows a heat map of the final geodesic distances for each voxel, and column 5 shows the associated Voronoi tessellation mapping each voxel to the geodesically nearest site. Figure~\ref{fig:process}, lower right, illustrates the LOS vs non-LOS voxels and pathways. 

\subsubsection{Geodesically Weighted Voronoi Site Update.} 

The next component of the algorithm is the updating of the site positions based on Equation~\ref{eq:2}. This is illustrated in Figure~\ref{fig:steps} row 2, column 4, and Figure~\ref{fig:process}. In a GPU kernel parallelized over voxels, we sum the moments $m_v\mathbf{x}_v$ of each \textit{line-of-sight} voxel in each Voronoi region. The Cuda atomic add operator is used for the summation. The geodesic weighting comes into play as the line of sight voxels which are path nodes that connect voxels which are around corners are weighted by the amount of voxels that they are sources of.

Next, a GPU kernel parallelized over the Voronoi sites is invoked, and the resulting site's sum of moments from the previous stage is normalized by the net mass to give the vector $\mathbf{x}_c$ from Equation~\ref{eq:2}, which is the new target position for the $\mathbf{x}_s$. We then cast a ray from $\mathbf{x}_s$ to $\mathbf{x}_c$ and move $\mathbf{x}_s$ only as far as possible before crossing the boundary.

\subsubsection{Data Reduction and Structured Data Layout.} 
\label{sec:layout}

Oftentimes, flow features of interest will comprise only a small portion of the full data. In this case, it makes sense to summarize and/or extract only the data within the level set restrictions of our tessellation as a way to reduce the data. Our method can optionally extract the raw data and couple it with the statistical summarization in a structured data layout before saving on disk. The data structure is hierarchically ordered by level set and connected components as shown in Figure~\ref{fig:hierarchy}. This ensures the components and layers are directly indexable and contiguous in storage, so one can directly load only the feature of interest for an efficient out of core approach. The statistical summaries, which are lighter weight, can be loaded first, explored, and then other data for selected features of interest can be loaded selectively.

The reduction from discarding the regions beyond the isobands can be approximated as $(n-r)\times (m+1) + n_l + n_c$ where $n$ is the number of volume elements in the full domain, $r$ is the size of the subset within the levels, and $m$ is the number of variables per volume element. $n_l$ and $n_c$ are the number of layers and components, respectively (for storing the offsets) which should be very small relative to $r$. Thus, when $(n-r)$ is large and or $m$ is large, the additional cost of storing coordinates becomes nearly negligible. In extreme-scale combustion simulations, $m$ can be greater than $100$, in which case the coordinates can be as little as $1\%$ of the total footprint of the reduced data. 

\vspace{2mm}
\begin{figure}
  \centering
  \includegraphics[width=0.99\linewidth]{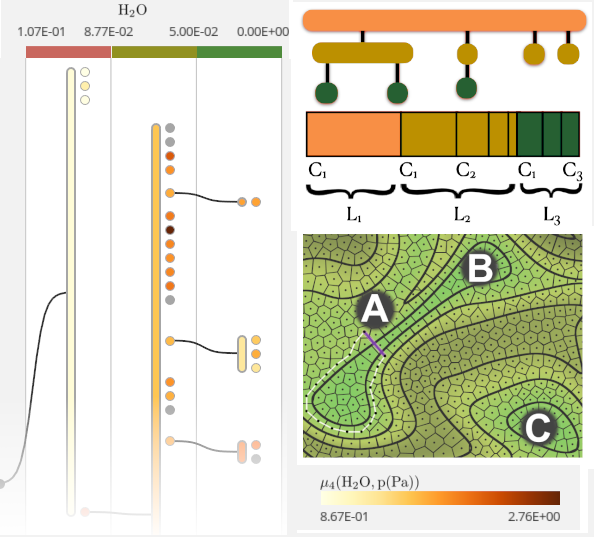}
  \vspace{-0.2in}
  \caption{Top Right: The data layout is sorted hierarchically by layer and component. Middle right: (A) White line is a rough measure of geodesic distance based on paths along Voronoi cells following the level set restriction; blue line is Euclidean distance. (B) and (C) are two separate connected components of the same layer. Left: The tree view in our visualization system showing the iso-values defining the layers and the connected components as nodes. The grey nodes are designated as components that are too small to compare with statistical measures.}
  \vspace{-4mm}
  \label{fig:hierarchy}
\end{figure}


\subsection{Role and Choice of Parameters}
\label{sec:params}

\textbf{Tessellation}. The tessellation is both dynamic and data dependent, and also depends on a few important parameters. First, one must have the scalar field to use for the level set restrictions. For example, in our combustion application we may choose temperature isotherms, or we may use the distance function from the flame surface. The set of isovalues must then also be chosen. If the method is used in situ, one must know in advance what kinds of value ranges the simulation will produce, and what isobands will be of interest for representing the flow features. In addition, if the tessellation is weighted based on a second scalar field, then this adds another aspect to what must be known in advance to appropriately. The parameter $\gamma$ controls the attraction force is equivalent to applying a non-linear re-scaling of the second scalar field used for weighting, and the effect it has depends on the data. The site density parameter $\alpha$ represents the choice of the number ratio of Voronoi sites to voxels. So this parameter can reasonably be decided in advanced with limited knowledge.

\textbf{Aggregation}. When aggregating data for spatial statistical summarization, information is lost if the raw data is not also output. Additionally, statistical aggregation is usually not a competitive form of pure compression, so if the primary goal is a reduced form of data to reconstruct the original data from, then this is probably not the best approach. Instead, we consider the statistical summarization as an explorable set of analysis results, which is useful for summarizing and analyzing certain aspects of the data, especially the dynamic spatial multivariate correlation structure. In this paper, we assume existing methods for statistical summarization of the segments in the tessellation will be used (\textit{e.g.} histograms, Gaussian's, Gaussian mixtures, etc.). It also depends on the method and application what the value of the summarization will be, how much space it will take, and how much of the raw data if any can be replaced with the summary. Our view is that the summaries have multiple opportunities for added value. First, they are lightweight and, when coupled appropriately with a visualization system, can support scalable interactive visualization of a larger domain than would otherwise be possible from using the raw data directly. Second, even if the summary is not used to replace the raw data, it represents an analysis result that would be useful anyways in addition to the raw data. And since simulations are already commonly unable to output full raw data at each time step due to I/O and storage limitations, outputting the more lightweight summaries at higher temporal fidelity would add value to the simulation result without reducing any information at only a nominal extra cost. The lighter weight summaries can be possible to stream through a network for in situ monitoring. Finally, considering the hierarchically structured data layout that our tessellation provides, the summaries corresponding to flow features can be extracted out of core, and if coupled with raw data, if it exists, can also be a basis for drilling down into the full details from the raw data. }

\section{Interactive Visualization System}
\label{sec:vis}

To showcase our feature-based approach, we designed and implemented a prototype visualization system. The system represents one promising way that the dynamic spatial decomposition approach could be leveraged. The system demonstrates the qualitative value of the approach, and we hope it can inspire more adaptations and ideas for similar work. 

The visualization workflow follows the natural hierarchical pattern for sub-selection that is implicit in spatial decomposition. As the user drills down into details, based on layers, connected components, Voronoi cells, smaller subsets of volume elements are brought into play for spatial statistical analysis. The layers, components, Voronoi sites, and volume elements are interacted on and visualized in 3 associated sets of views as shown in Figure.~\ref{fig:teaser}.  
Since the system supports interactive multivariate statistical analysis, data from many fields may need to be active simultaneously. The hierarchical sub-selection becomes useful in this case, not only as a means to narrow down the features of interest, but also to improve performance since sub-selections at each level will significantly reduce the amount of active data.

\begin{figure*}[ht]
  \centering
  \includegraphics[width=0.99\linewidth]{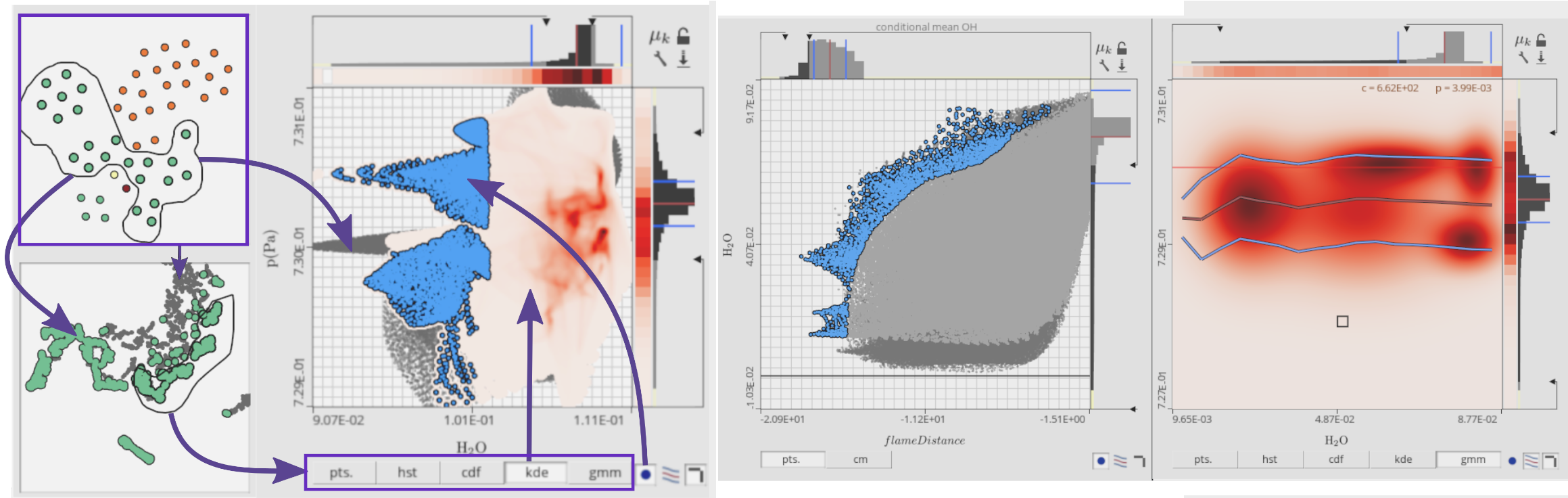}
  \vspace{-0.12in}
  \caption{Upper left) t-SNE projection of components with lasso selection. Lower left) projection of Voronoi regions within selected components with second level lasso selection. Middle Left) KDE plot with voxel sub-selection (blue), and sub-marginals displayed. Middle right) scatter plot option is selected. Right) GMM with 1D conditional plot on top.}
  \label{fig:jointPlot}
  \vspace{-0.2in}
\end{figure*}

{
\subsection{Motivation and Design Goals}

This system in its first iteration was designed based on discussions between visualization researchers and combustion scientists. Both the visualization researchers and scientists are co-authors of this paper. Since, to our knowledge, no other systems exist that could explore the same data in sufficiently comparable way, and due to the limited length and scope of the paper, we are not able to do a thorough presentation of the system and robust evaluation at this stage. Instead we view this system as a starting point, and proof of concept for how to utilize the tessellation result, which may be built on, optimized, and evaluated in depth in future work. 
We provide below the set of visualization and analysis goals used to inform our design. In Section~\ref{sec:discussion} we further lay out our vision for improvement and evaluation of the system.

\textbf{Design Goals:} 
\begin{enumerate}[noitemsep,topsep=0pt]
    \item Be able to use the system on both raw and aggregated data for testing and comparison.
    \item Be able to link and visualize the multivariate statistical plots and the features in the spatial domain.
    \item Support the following common statistical plots: 1/2D probability density functions (PDF's), cumulative distribution functions (CDF's), 1/2D conditional plots, and scatter plots.
    \item Be able to compare different candidate models for the probability distribution functions.
    \item Be able to inspect the higher order joint statistical moments and how well the PDF models preserve them.
    \item Be able to drill down into the details with a focus on the hierarchical structure of the data that is decomposed around flow features.
\end{enumerate}
}

\subsection{Isobands and Connected Components} 

The first set of views that the user will interact with are for selecting isobands and their connected components. A tree-based visualization is used to show the nested hierarchy as depicted in Figure.~\ref{fig:hierarchy}, and another linked view represents the components based on a 2D t-distributed stochastic neighbor embedding (t-SNE) projection, as can be seen in Figure.~\ref{fig:jointPlot} (Left). The input to the projection is a featurization of the components, for example based on the bin values of histograms, parameters of parametric statistical models, or statistical moments. t-SNE will try to group the points with similar statistical features together. The points can be selected by the user through a lasso selection tool in the projection or from the tree view. The lasso selection operation has several modes besides new selection ($\circ$), based on set combination operations with the existing selection ($\cap, \cup,$ or $\setminus$).

 In the projection, the points are color coded by isoband, but in the tree view, that information is implicitly encoded by level of the node. Thus, for tree nodes, color encodes other information. For example, in Figure.~\ref{fig:hierarchy} color is used to encode $\mu_4(\ch{H_2O},p)$, which is the cokurtosis of hydrogen mixture fraction and pressure. Since very small components have too few voxels to get quality statistics, we color encode them in gray instead. This can also be helpful information since those small components may be particularly meaningful ({\it e.g. } in combustion they may indicate the incipient development of an ignition event).

\subsection{Voronoi Cell Views}

The Voronoi regions are also projected into t-SNE projection views. Figure.~\ref{fig:jointPlot} (lower left) shows such a projection. After selecting a set of components, the Voronoi cells in those components are highlighted in the Voronoi region projections and become selectable using the same lasso tool with the same set operator modes. 

In addition to t-SNE projection based on statistical features, we found that a projection based on spatial similarity is a useful option for one or more of the linked Voronoi region projections, since it can group points based on spatial proximity. In Figure~\ref{fig:teaser} (right) the bottom projection is based on such a 2D spatial projection. For this projection, we use a pre-computed distance/similarity matrix based on a custom pairwise distance measure between Voronoi sites. Since the tessellation is restricted by level sets, it makes sense to utilize the geodesic/path distance obeying the restrictions, as illustrated in Figure.~\ref{fig:hierarchy} (A). The blue line shows the Euclidean distance, while the white line shows geodesic distance. An example where this useful in capturing the condition where regions around a folded surfaces (\textit{e.g.} flame) are geodesically distant but yet close together in Euclidean distance. In combustion, for example, this may preclude or represent interesting interactions of the flame as the Euclidean distance approaches the laminar flame thickness while the geodesic distance remains larger. One example distance function that can be used is,
\vspace{-2.0mm}
\begin{equation}
d(s_i, s_j) = \left( \frac{||\mathbf{x}(s_i) - \mathbf{x}(s_j)||_2}{ gd(s_i, s_j)} \right) \textbf{\textit{ if } } P_{i,j} \text{ exists } \textbf{\textit{ else }} c
\end{equation}
\noindent
where $P_{i,j}$ is the graph geodesic of adjacent Voronoi sites, $gd$ is the path's distance, $c \geq 1$ is a parameter where a higher value results in more separation of sites outside of the same component. To find $P_{i,j}$, we compute the connectivity of the Voronoi sites based on region adjacency, and then Johnson's all-pairs shortest path algorithm (which internally uses Dijkstra's algorithm) is used. Since the neighborhood is also restricted to within connected components the computation is parallelized by letting different threads operate on different components. The  Boost graph library's implementation is used, which has time complexity $O(|V|\cdot |E| \cdot log (|V|))$, where $V$ is the set of nodes, and $E$ is the set of edges. An alternative to Johnson's/Dijkstra's is to use the Floyd-Warshall all pairs shortest path algorithm which has time complexity $O(|V|^3)$. Since our graph is only locally connected it is sparse enough that $O(|V| \cdot |E| \cdot log(|V|))$ is better than $O(|V|^3)$.   

\subsection{Interactive Joint Plots}

Once a selection of Voronoi regions is made, statistical plots are generated, either from their pre-aggregated statistics, or from the raw data loaded out of core based on the aforementioned structured data layout. 

Based on the experience of the domain scientists, the following statistical plots are chosen: joint PDF's and their marginals, joint CDFs, conditional plots (which estimate one variable conditioned on 1 or 2 others), and scatter plots. Since each of these plots share the same axis with the same scaling, we can switch between them, or combine layers on a single interactive generalized joint plot. 

The joint plot includes information from both the background layer which is generated from the set of selected components, and foreground layer which is generated from the set of Voronoi regions within the components. The background layer in the joint plot is a static gray scatter plot, while the foreground layer is plotted over the background and depends on the selected plot mode. There are also marginals for both the background layer which are shown as gray histograms drawn as bar charts, and marginals of the foreground data as heatmaps beneath the bar charts as seen in Figure~\ref{fig:jointPlot}. The user can zoom and pan by scrolling or dragging the mouse when within either the joint plot (to zoom/pan both axis), or over one of the marginals (to zoom/pan only one axis). Highlighting on the background marginals, and arrows are used as a form of mini-map to depict the zoom level.

The joint plot of the foreground layer can be either scatter plot, histogram, CDF, kernel density estimation (KDE), or Gaussian mixture models (GMM's). The KDE plots are useful for capturing high-level detail and smoothness in the potentially complex nonlinear distributions. Histograms and GMM's are models which are commonly used for in situ statistical summarization. By using our system on existing data, before employing in situ, the choices of models and their parameters can be tested and the user can get an idea what information might be lost or misrepresented using the candidate methods. As the user interacts with the data, the PDF's may be interactively recomputed as the data inputs and the data range changes. Since the KDE and GMM plots are often not able to be recomputed immediately at interactive  rates, the plot will automatically switch from KDE or GMM to a histogram, for example, while zooming/panning is occurring.

To evaluate the PDF's, the system estimates how well they preserve joint moments, which are used in the analysis of turbulent flows \cite{sreeni97,bennett2011}. The moments ({\it i.e. } mean, covariance, coskewness, and cokurtosis) estimated from the PDF's are compared against the moments computed from the raw data. Each plot has a button (in the upper right corner, labeled $\mu_k$) that can be pressed to open/close an attached table that is auto-generated as a LaTeX document as shown in Figure.~\ref{fig:teaser}. This functionality can also aid featurization of high-order moments-based anomaly detection algorithms to \cite{aditya2019,kolla2020}.

If voxel data is available, the voxels deriving the foreground can be selected using the lasso tool and aforementioned set combination operators. The selections are linked across views and plotted as blue points. This layer can be toggled on/off per-plot to manage occlusion. These selections are also linked in the 3D view where their bounding surfaces are rendered. Conditional plots can be toggled on/off as well, and are rendered on the top most layer as a red curve estimating mean $y$ as a function of $x$, with blue lines above and below the curve at $\pm 1$ standard deviation. 2D conditional plots, showing estimated mean $z$ as a function of $(x,y)$, rendered as a heatmap, is also supported. For these plots, the forground layer can also be toggled to show a scatter plot of $(x,y)$ as shown in Figure~\ref{fig:jointPlot} (second from right).

\begin{figure}[h]
    \centering
    \includegraphics[width=1.0\linewidth]{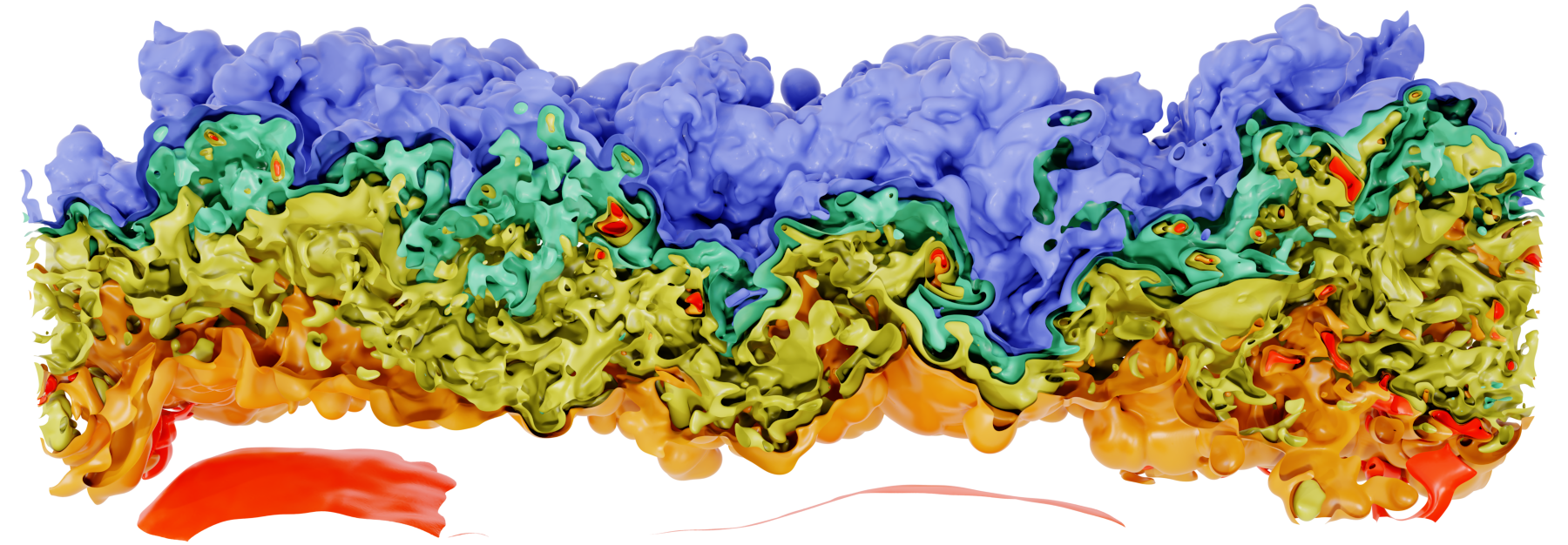}
    \vspace{-0.22in}
    \caption{Isotherms in the premixed hydrogen/ammonia/nitrogen flame.}
\vspace{-4mm}
    \label{fig:ammonia1}
\end{figure}

\section{Performance}

The data used to obtain these results is the hydrogen/ammonia/nitrogen flame data described in Section~\ref{sec:case}. Five isobands of temperature were used. The bands are highly contorted as can be seen in Figure~\ref{fig:ammonia1}. This makes the data a good stress test for the algorithm as the Voronoi sites need to find their way around complex obstacles.

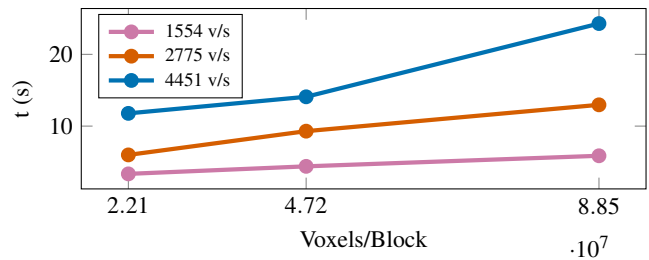
\begin{figure}[h]
\centering
\begin{tikzpicture}[scale=1.0]
    \definecolor{c1}{RGB}{204,121,167}
    \definecolor{c2}{RGB}{213,94,0}
    \definecolor{c3}{RGB}{0,114,178}
    \definecolor{c4}{RGB}{0,158,115}
    \definecolor{c5}{RGB}{0,0,0}    
    \definecolor{c6}{RGB}{230,159,0}        
    \begin{axis}[
        legend style={nodes={scale=0.8, transform shape}},
        height=2.4cm,
        width=.85\linewidth,
        scale only axis,
        ylabel near ticks, 
        xtick=data,
        legend pos=north west,
        xlabel=Voxels/Block,
        ylabel=t (s),
        every axis plot/.append style={ultra thick}
    ],
\addplot[color=c1,mark=*] plot coordinates {
 (22118400, 3.343)
(47185920, 4.392)
(88473600, 5.864)};
\addplot[color=c2,mark=*] plot coordinates {
 (22118400, 5.986)
(47185920, 9.294)
(88473600, 12.964)};
\addplot[color=c3,mark=*] plot coordinates {
(22118400, 11.774)
(47185920, 14.082)
(88473600, 24.289)};
    \legend{~1554 v/s, ~2775 v/s, ~4451 v/s}
    \end{axis}
    \end{tikzpicture} 
     \vspace{-0.10in}
 \caption{Computation time per-node for convergence to within a mean $d_s$ of $0.5$ (unit voxel length). Each line represents a different block size for distributed computation. Since the computation is independent for each block of data in a distributed setting, the scalability can be assumed to depend primarily on the number of GPU nodes available.}
 \label{fig:scaling}
\end{figure}

First, we evaluate the computation time with respect to site density and number of voxels on a node; shown in Figure~\ref{fig:scaling}. The performance drops as the the number of voxels on a GPU node increases. The primary bottleneck is the Voronoi classification described in Section~\ref{sec:VT}. This computation is parallelized on the GPU using one thread per voxel without communication. Thus we can expect the GPU parallelism to scale well with the number of GPU threads. However, the performance decreases approximately linearly with respect to site density, as lower site density means longer paths traced out per voxel/thread.

\begin{figure}[h]
\centering
\begin{tikzpicture}[scale=1.0]
    \definecolor{c1}{RGB}{204,121,167}
    \definecolor{c2}{RGB}{213,94,0}
    \definecolor{c3}{RGB}{0,114,178}
    \definecolor{c4}{RGB}{0,158,115}
    \definecolor{c5}{RGB}{0,0,0}    
    \definecolor{c6}{RGB}{230,159,0}        
    \begin{axis}[
        legend style={nodes={scale=0.8, transform shape}},
        height=2.0cm,
        width=.85\linewidth,
        scale only axis,
        ylabel near ticks, 
        xtick={1,10,20,30,40, 50},
        legend pos=north east,
        xlabel=Number of Updates,
        ylabel=Mean $d_s$,
        every axis plot/.append style={ultra thick}
    ],
\addplot[color=c1] plot coordinates {
( 1,  2.70571893 )
( 2,  1.0961861  )
( 3,  0.68749321 )
( 4,  0.50765499 )
( 5,  0.39981147 )
( 6,  0.33276758 )
( 7,  0.28887818 )
( 8,  0.25274347 )
( 9,  0.22452515 )
( 10, 0.20284192 )
( 11, 0.18575097 )
( 12, 0.17084857 )
( 13, 0.15702671 )
( 14, 0.14666087 )
( 15, 0.13712513 )
( 16, 0.12825877 )
( 17, 0.11906909 )
( 18, 0.11180129 )
( 19, 0.10710157 )
( 20, 0.10212681 )
( 21, 0.09670773 )
( 22, 0.09261008 )
( 23, 0.08879266 )
( 24, 0.08569172 )
( 25, 0.08228326 )
( 26, 0.08013469 )
( 27, 0.0767919  )
( 28, 0.07506164 )
( 29, 0.07216286 )
( 30, 0.06934162 )
( 31, 0.06680832 )
( 32, 0.06558862 )
( 33, 0.06458174 )
( 34, 0.06333005 )
( 35, 0.06084203 )
( 36, 0.05879263 )
( 37, 0.05793648 )
( 38, 0.05703287 )
( 39, 0.05566479 )
( 40, 0.05365745 )
( 41, 0.05276975 )
( 42, 0.0512602  )
( 43, 0.0512692  )
( 44, 0.05068514 )
( 45, 0.04897606 )
( 46, 0.04657789 )
( 47, 0.04541899 )
( 48, 0.04477949 )
( 49, 0.04361152 )
( 50, 0.04370573 ) };
\addplot[color=c2] plot coordinates {
( 1,  2.67872685 )
( 2,  1.11174664 )
( 3,  0.6885432  )
( 4,  0.49929662 )
( 5,  0.3942369  )
( 6,  0.32657205)
( 7,  0.28045714)
( 8,  0.24598102 )
( 9,  0.21913285 )
( 10, 0.19627841 )
( 11, 0.17869736 )
( 12, 0.16437089)
( 13, 0.15186591 )
( 14, 0.14165621 )
( 15, 0.13283664 )
( 16, 0.12509265 )
( 17, 0.1177685  )
( 18, 0.11113545)
( 19, 0.10553779 )
( 20, 0.10039299 )
( 21, 0.09523247 )
( 22, 0.09049684 )
( 23, 0.08639474 )
( 24, 0.08297912)
( 25, 0.0798249  )
( 26, 0.07700205 )
( 27, 0.07428027 )
( 28, 0.07176226 )
( 29, 0.06944073 )
( 30, 0.06754137)
( 31, 0.06533839 )
( 32, 0.06361158 )
( 33, 0.06171412 )
( 34, 0.06007918 )
( 35, 0.05832661 )
( 36, 0.05692681)
( 37, 0.05541249 )
( 38, 0.05429122 )
( 39, 0.05299556 )
( 40, 0.05178331 )
( 41, 0.05048687 )
( 42, 0.04990645)
( 43, 0.04868009 )
( 44, 0.0474733  )
( 45, 0.04605379 )
( 46, 0.04483251 )
( 47, 0.04368965 )
( 48, 0.04270802)
( 49, 0.04202819 )
( 50, 0.04164374) };
\addplot[color=c3] plot coordinates {
( 1,  2.39015704  )
( 2,  1.01728894  )
( 3,  0.6307451   )
( 4,  0.45813613  )
( 5,  0.36114222  )
( 6,  0.29859493)
( 7,   0.25522687 )
( 8,   0.22311439  )
( 9,   0.19838243  )
( 10,  0.17890165  )
( 11,  0.16319309  )
( 12,  0.15019132)
( 13,  0.13897928  )
( 14,  0.12959173  )
( 15,  0.12118988  )
( 16,  0.11394664  )
( 17,  0.10699982  )
( 18,  0.10153641)
( 19,  0.09655398  )
( 20,  0.09193093  )
( 21,  0.08754708  )
( 22,  0.08344697  )
( 23,  0.07992134  )
( 24,  0.07700163)
( 25,  0.07339588  )
( 26,  0.07059946  )
( 27,  0.06792262  )
( 28,  0.06566141  )
( 29,  0.06331091  )
( 30,  0.06112617)
( 31,  0.05913134  )
( 32,  0.0571855   )
( 33,  0.05533534  )
( 34,  0.05372693  )
( 35,  0.05222514  )
( 36,  0.05063066)
( 37,  0.04925872  )
( 38,  0.04795452  )
( 39,  0.04681903  )
( 40,  0.04555316  )
( 41,  0.04440904  )
( 42,  0.04340931)
( 43,  0.04266628  )
( 44,  0.0412739   )
( 45,  0.04036446  )
( 46,  0.03941522  )
( 47,  0.03841206  )
( 48,  0.03755584)
( 49,  0.03698712  )
( 50,  0.03631224)
    };
    \legend{~7142 v/s, ~3571 v/s, ~1786 v/s}
    \end{axis}
    \end{tikzpicture} 
     \vspace{-0.1in}
 \caption{The convergence of mean $d_s$ over the number of geodesically weighted centroidal update steps, shown for 3 different site densities. The algorithm converges similarly in each case to low mean $d_s$ (less than $0.5$ voxel widths), very rapidly. The  continues to improve slightly over time with diminishing returns on the benefit for the cost.}
 \label{fig:convergence}
\end{figure}
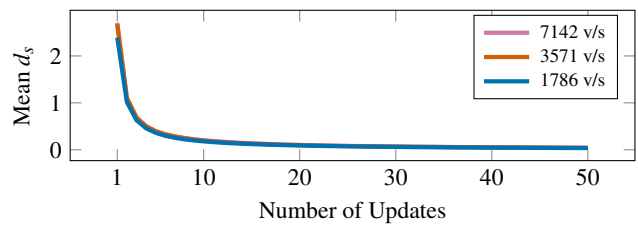

The second result, in Figure~\ref{fig:convergence}, shows how quickly the tessellation converged in terms of steps in Algorithm 1. Diminishing returns for the computation cost are seen after only a handful iterations. Since the computation is independent over each node/block of the block decomposition domain, we were able to use a single user machine to perform these first two tests; an Arch-Linux machine with an Intel(R) Core(TM) \verb|i7-5930K CPU @ 3.50GHz, 64 GB| of main memory, and an \verb|NVIDIA GeForce GTX TITAN X| which has 12 GB of memory. The number of Cuda threads per-block was set at 128, which gave us the best performance. The Cuda code was compiled with compiler flags, `nvcc -std=c++11 -O3 -w -m 64 -rdc=false -ftz=true -use\_fast\_math -Xptxas -O4, -Xcompiler -O4 -arch=sm\_52'. 

\begin{table}[H]
  \label{performance}
  \scriptsize%
	\centering%
  \begin{tabu}{%
	r%
	*{6}{c}%
	*{2}{r}%
	}
  \toprule
  \textbf{Nodes}   &   8     &     16 &    32 &    48 &   60  \\
  \textbf{GPUs}    &  48     &     96 &   192 &   288 &  360  \\
  \textbf{time}(s) &  26.939 & 12.996 & 5.618 & 3.990 & 3.091 \\  
  \bottomrule
  \end{tabu}%
  \vspace{0.15in}
    \caption{Parallel Performance results for Ammonia Data of Size $3456\times1280\times2560$. The computation was performed till mean $d_s$ reached $0.25$ voxel lengths.}
\end{table}
\vspace{-0.19in}

Last, we performed a full distributed parallel test for the combustion data on the Summit supercomputer. The volume has a size of $3456 \times 1280 \times 2560$ voxels. The site density was ~2700 voxels per Voronoi site. We ran the system until it reached the point where mean $d_s$ dropped below $0.25$ unit voxel widths. The results are shown in Table 1. Each of the nodes on Summit has 6 NVIDIA V100 Tesla accelerators, 512 GB of DDR4 memory, and two IBM POWER9 CPUs. We ran our code with 1 GPU and 1 CPU per rank, and 6 ranks per node. We testing the performance using 48, 96, 192, 288, and 360 GPUs. Since Summit has 6 GPUs per-node, we mapped this to 8, 16, 32, 48, and 60 nodes respectively. The run command for example was `jsrun -n48 -a1 -c4 -g1 ./t8' for the case of 48 GPUs. The CUDA code was compiled with NVCC, with command-line arguments: \verb|-std=c++11| \verb|-O3| \verb|-w| \verb|-m 64| \verb|-rdc=true| \verb|-Xptxas| \verb|-O3|. The CUDA version was 11.4.2. The C++ program that links the GPU code was compiled with GCC 10.2.0 (the GNU Compiler Collection) with arguments: \verb|-std=c++11 -O3| \verb|-fpic| \verb|-fopenmp|. Since each node took a different amount of time to complete, we report the worst case, which represents the overall time for the entire data to process. The CUDA kernels used a block size of 128. 




\section{Applications}
\label{sec:case}

\begin{figure}[H]
    \centering
    \includegraphics[width=0.99\linewidth]{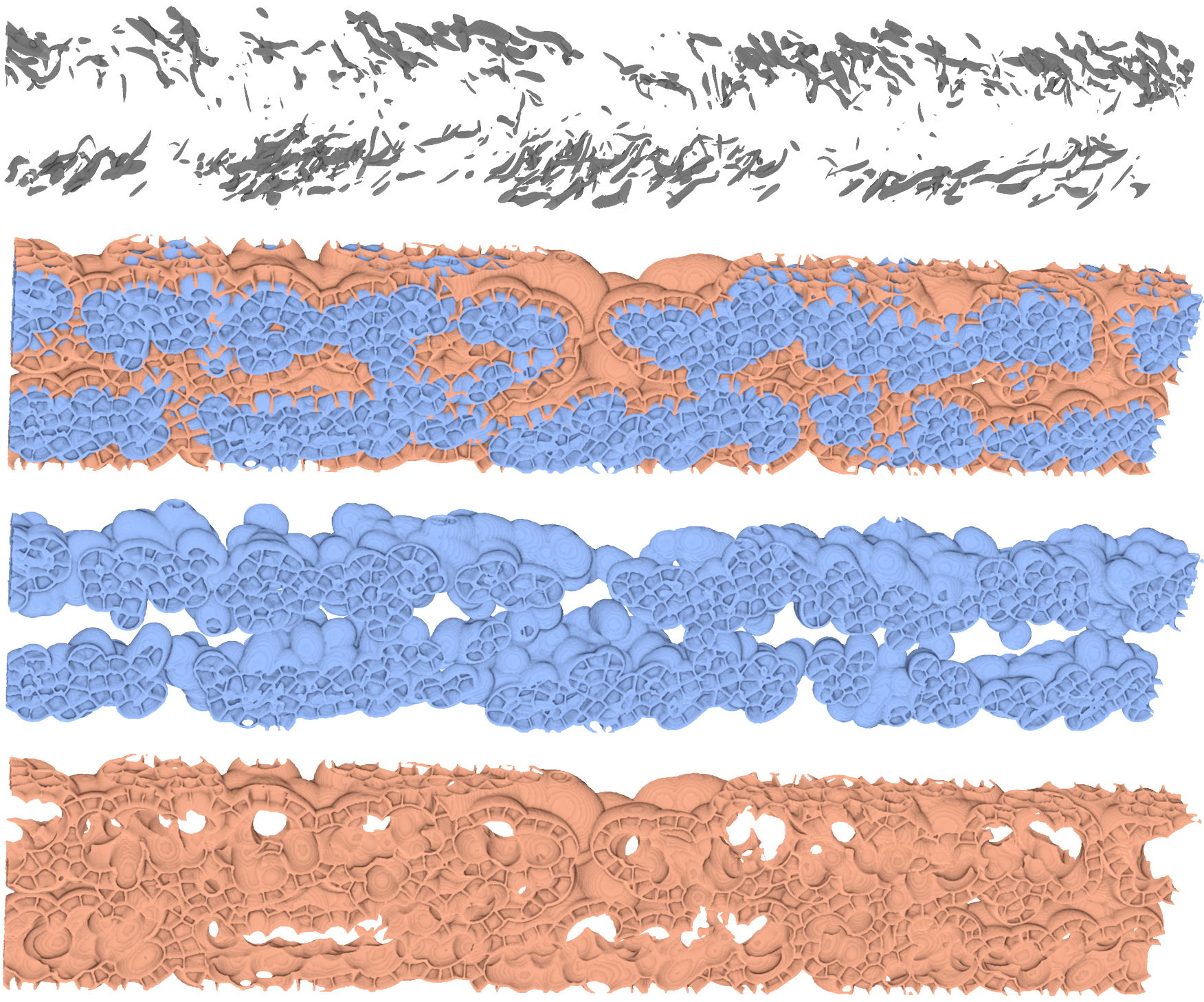}
    \vspace{-0.1in}
    \caption{Channel flow decomposition using the $\lambda_2$ field to encapsulate vortical structures, and the distance function to define iso-levels. The dark grey surface represents the vortical structures.}
    \vspace{-2mm}
    \label{fig:lambda2}
\end{figure}

We tested our algorithms on two types of flow data, turbulent combustion and wall bounded channel flow. The flame surface in the turbulent combustion data is more connected, but still highly complex, while the channel flow data has many small vortical structures. The combustion data features strong coupling between the flame wrinkling and the underlying turbulent strain, where the flame response occurs in a composition phase space (dependent locally on species concentrations and temperature which control the burning rate and are modulated by the turbulent strain), while turbulent strain evolves in physical space. Hence, there is an inherent need to interactively toggle between physical and composition space to drill down and understand causality. Depictions of the two datasets and tessellations made using them are shown in Figure~\ref{fig:ammonia1} and Figure~\ref{fig:lambda2}, respectively.


\subsection{Hydrogen/Ammonia/Nitrogen Turbulent Flame}

We apply the nested hierarchical approach presented here to turbulent combustion data from Direct Numerical Simulation (DNS) of a premixed hydrogen/ammonia/nitrogen flame \cite{wiseman21}. The DNS study aims to gain an understanding of the combustion behavior of hydrogen/ammonia/nitrogen as a carbon-free replacement for conventional fuel ({\it i.e.}, natural gas). The simulation was run on 900 nodes of the Oak Ridge National Laboratory OLCF Summit cluster requiring approximately 110k node-hours.   The aggregate amount of data produced comprises 400 checkpoint files, where each checkpoint file is 2TB. 

Due to the nature of turbulent combustion and the need to resolve all flow and chemical scales, turbulent combustion DNS data is characterized by a large separation of scales, with portions of the data often not relevant for statistical analysis or over-resolved on the numerical grid that is designed to capture the smallest scales (\textit{e.g.}, regions of hot products feature small gradients/large flow and chemical features). The Voronoi tessellation effectively reduces the amount of data to be processed for the purpose of statistical analysis and allows for efficient and fast data exploration. While conventional statistical analysis ({\it e.g.}, (joint) statistical distributions, statistical moments, correlations between species and temperature and between reaction rates and strain rate) is important in turbulent combustion, the analysis of spatially and temporally intermittent features is often of interest ({\it e.g.}, flame-flame interactions, reactant pocket formation enriched with minor species that are chemically crucial, and formation of localized flame extinction holes). The hierarchical selection of connected components together with distance-based metrics, and with the linked views between spatial and statistical features, allows scientists to quickly identify and select specific flame features and directly extract necessary statistics to understand causality between the `turbulence-chemistry' interactions and to steer further downstream analysis, and adaptive labeling of `feature' data for searchable training and validation for machine learning.

The mass fraction of H$_\mathrm{2}$O, normalized by the values obtained in burned and unburned gas of a laminar, unperturbed flame at the same conditions as the turbulent flame, acts as a so-called progress variable, which monotonically increases from 0 to 1 across the flame brush. This progress variable is usually used to identify specific zones of a premixed flame and specific values can be selected to obtain iso-surfaces that identify the various zones in the flame front ({\it i.e.} preheat zone, reaction zone, oxidation zone in a premixed flame). Here, we choose a specific value that represents the reaction zone, {\it i.e.}, the location where the heat released by the flame is at its peak. We refer to the iso-surface corresponding to this value as the flame surface. Of particular interest is the analysis of the value of other quantities on the flame surface. For example, a large value of the OH radical species represents a region that is burning particularly strongly and in the present dataset, this corresponds to a flame element that experiences preferential diffusion effects (\textit{i.e.}, hydrogen reactant diffusing faster than heat), where the burning is amplified locally by a large influx of fuel. In other regions of the flame, OH is low, which represents flame elements that are locally extinguishing, {\it i.e.}, strain and stretch-induced by turbulence disrupt the reactions to a large extent. Distance-based metrics also provide a conditioning variable in physical space indicating the response of different zones in the flame, upstream and downstream of the flame isosurface, to the strain rate.  Distance function iso-contours represent the locus of all points that are equidistant from the flame surface.

\subsection{Turbulent Channel Flow at High Reynolds 
Number}

Here we examine is the incompressible turbulent channel flows at a high Reynolds number, $Re$ (Figure~\ref{fig:lambda2}). High $Re$ wall-bounded turbulence has great importance in many engineering applications with moving objects at high speed. However, the fundamental dynamics of wall-bounded flows are not yet fully understood. The turbulent channel flow is one of the simplest canonical flows to study the high $Re$ wall-bounded turbulence. Still, the DNS of high $Re$ turbulent channel flows is quite expensive due to the scale separation similar to the turbulent combustion as explained above. The simulation performed with 33K nodes in Mira (Bluegene/Q, ALCF) used 242 billion DOFs, and it remains the largest DNS simulation of wall-bounded turbulence  \cite{Lee2013,Lee2014,Lee2015}. 

In this study, we use the $\lambda_2$ field, which is calculated from a subset of the aforementioned DNS. The interaction of coherent structures in turbulent flows is a long-standing problem. Naturally, the methods to identify vortex structures are well developed. $\lambda_2$ is one of the widely used vortex identification methods. It uses the eigenvalues of a $3\times3$ matrix constructed with a velocity gradient tensor\cite{Jeong1995}. Even though the vortex identification method is mature itself, using it to study the dynamics of high $Re$ wall-bounded turbulence is still a challenging task because the identified vortex structures are too numerous to analyze readily from the raw data\cite{Lozano-Duran2014a}. Note that other types of high $Re$ turbulence also encounter the same flow complexity. The dynamic nested hierarchical statistical decomposition approach can efficiently identify the vortical structures and analyze their interactions with other state variables, such as velocity and pressure fields.




\section{Discussion} 
\label{sec:discussion}






\noindent
{\bf Block Based Distributed Parallelism.}
For large-scale domains, a distributed approach is needed. There are two possibilities; (1) perform the CVT globally over the whole domain, which will be computationally intensive due to the communication requirements across nodes, or (2) compute the CVT separately on large blocks of the domain on different ranks. We have only implemented the second option. The result is that the boundaries of the blocks will act as further restrictions so that the Voronoi sites will be aligned to those boundaries in addition to the isobands. Such alignments are already implicit at the boundaries of the full domain, and the alignment to the block boundaries does not break the level set based hierarchy, or alignment of the CVT to the level set surfaces. Additionally, it offers one advantage over a global CVT; whole blocks can be loaded without clipping the tessellation. This is practical because large datasets are often analyzed post-hoc through block-wise sub-selection due to the large scale. It also makes a more efficient implementation possible and is trivial to scale up. 

{\bf CVT Algorithm and Optimality.} In general, it is difficult to prove guarantees for CVTs~\cite{du2006convergence}. Liu et al. provide theory on CVT convergence~\cite{liu2009centroidal}. Despite theoretical challenges, CVT algorithms adaptively minimize energy and are argued to have practical expectations of convergence. For our algorithm it depends on the definition of $\mathbf{x}_{c}$, the new target site position in the update step for Loyde's algorithm. As site density increases and/or the restriction complexity decreases, $\mathbf{x}_c$ approaches the Euclidean centroid. While the tessellation will evolve from less optimal to more optimal distributions, we are not able to prove theoretical guarantees. A more optimal definition is also possible, but it is unclear what that should be in order to balance efficiency and quality. Our implementation can be extended or improved in the future by redefining $\mathbf{x}_{c}$ and modifying the update step. As an alternative to Loyde's algorithm, one can also try an approach based on limited memory BFGS method for large scale optimization (L-BFGS)~\cite{liu1989limited}. 



{\bf Need to Choose Parameters A Priori and Loss of Information.} As discussed in Section~\ref{sec:params}, there is a significant limitation in the need to specify parameters before hand, and the choice of parameters depends on dynamic data. This becomes a bigger concern when one is less clear what kind of data they can expect the simulation to produce. When one has some existing data which they expect to have similar characteristics to the data the simulation will produce, then they can apply the methods on the raw data and use our prototype system to explore the results while being able to compare what information may be lost from the raw data given a parameterization of the methods. As future work, one may explore robust dynamic methods to adapt the statistical aggregation/modelling approach as the data evolves based on the uncertainty or error. For example, when a model is no longer able to represent the data sufficiently, one can trigger more output of the raw data, or employ a more expensive yet superior algorithm or model.

{\bf Applicability and Extensibility.} 
Our implementation works on Cartesian grids and rectilinear grids. For rectilinear grids, one can perform the tessellation as if it were a Cartesian grid in which case the tessellation will be stretched to match the grid. In some cases, this approach is sensible because it results in more homogeneity in region sample sizes (volume elements per region), which may yield greater statistical consistency. However, one can also use the stretching function for the tessellation. The algorithm can be extended to work in most cases where a discrete binning of volume elements is acceptable.

{\bf Impact of the Visualization Approach} While visualization tools for 3D rendering and surface analysis of the large scale data have matured over the years, there is still a lack of scalable tools for exploring large scale multivariate data and spatial statistical correlation. Our work contributes new ideas and software to help push the state of the art in this area, and can be expanded on by future research.

{\bf Evaluation and Refinement of Visualization Design.} The visualization system and overall approach has clear motivation and has been applied to help combustion scientists better understand their data. The design is still in an early stage and represents only one for approach visualizing spatial statistics using the hierarchical segmentation. We have determined several ways it can be improve. First, a direct selection of connected components in the 3D view would be useful. This could be supported either based on an explorable image with depth buffers~\cite{tikhonova2010explorable}, through 3D ray intersection testing. Second, we would like to region growing of the selections based on spatial or statistical proximity. Lastly, the visualization system needs better support for temporal visualization. The statistical summaries of components can be efficiently visualized through time utilizing a dynamic tracking graph~\cite{lukasczyk2019dynamic} and the structured data layout from Section~\ref{sec:layout}. 

Once the visualization system has matured, a more robustly evaluation can be done. Evaluation of methods designed for use by qualified scientists for exploratory visualization and analysis can be challenging, because few are qualified to participate in studies, and appropriate tasks are difficult to define for visualizations when the result of the visualization is qualitative insight. If a user study is overly simplified or narrow in scope, it can give a false impression of the usefulness and potential of the design~\cite{greenberg2008usability}. Feedback from the participating scientists on the value of the methods and tools is included in Section~\ref{sec:case}. However, the qualified scientists we have worked with are co-authors and helped design the system around their interests, so there is a level of potential bias that may be expected in their direct feedback, and the potential use cases for the methodology spans domains we are not experts in. Our vision for the future evaluation and refinement of the system involves iterative refinement based on lessons learned as we apply it in our domain~\cite{8977377}, as well as through as much feedback as can be gained by other users outside our domain.

\section{Conclusion} 
\label{sec:future}

We believe that the workflow presented in this paper can help drive future scientific researchers closer to realizing the promises of ever increasing large and complex scientific data. 

In the near term, we envision semi-automated in situ workflows where the scientist needs to keep up with and play an active role in the process from simulation to analysis. The ability to identify statistical patterns and their characteristics manifested in both physical and high dimensional phase space coordinates will enable  `feature or event' based parameterization. As part of a holistic in situ dynamic workflow, and the detection of these features could trigger further statistical spatiotemporal analysis, `event' labeling and storage for data driven reduced-order surrogate modeling, adaptive mesh refinement, or adaptive in situ manifold learning as anomalous events occur.

In addition, our visualization software utilizing this methodology can be expanded to support a workflow for interactive statistical featurization, and event modelling, and then can automatically export the models into the target language and workflow model/system, to be rapidly deployed for into their scientific pipelines.  Ultimately, this would act as a coupling of the discovered spatial statistical pattern with a reproducible workflow that can be automated. This could then be applied in situ, \textit{e.g.} within a dynamic workflow for feature detection and subroutine triggering.

\acknowledgments{We thank Qi Wu for stimulating discussion. This research is sponsored in part by the US Department of Energy through grant DE-SC0019486. The work at Sandia National Laboratories was supported by the US Department of Energy, Advanced Scientific Computing Research Office. Sandia National Laboratories is a multimission laboratory managed and operated by National Technology and Engineering Solutions of Sandia, LLC., a wholly owned subsidiary of Honeywell International, Inc., for the U.S. Department of Energy’s National Nuclear Security Administration under contract DE-NA0003525. The work at Indian Institute of Science was supported by the Arcot Ramachandran Young Investigator Award provided to Konduri Aditya. An award of computer time for the turbulent channel flow dataset was provided by the Innovative and Novel Computational Impact on Theory and Experiment (INCITE) program. This research used resources of the Argonne Leadership Computing Facility, which is a DOE Office of Science User Facility supported under Contract DE-AC02-06CH11357.}

\bibliographystyle{abbrv-doi}

\bibliography{00-Main.bib}
\end{document}